\documentclass[10pt,letterpaper]{article}
\usepackage[top=0.85in,left=2.75in,footskip=0.75in]{geometry}

\usepackage{amsmath,amssymb}

\usepackage{changepage}

\usepackage[T1]{fontenc}

\usepackage[utf8x]{inputenc}

\usepackage{textcomp,marvosym}

\usepackage{cite}

\usepackage{nameref,hyperref}

\usepackage{microtype}
\DisableLigatures[f]{encoding = *, family = * }

\usepackage[table]{xcolor}

\usepackage{algorithm}
\usepackage{algorithmic}
\usepackage{url}

\usepackage{array}

\newcolumntype{+}{!{\vrule width 2pt}}

\newlength\savedwidth

\newcommand\thickhline{\noalign{\global\savedwidth\arrayrulewidth\global\arrayrulewidth 2pt}%
\hline
\noalign{\global\arrayrulewidth\savedwidth}}

\raggedright
\usepackage[aboveskip=1pt,labelfont=bf,labelsep=period,justification=raggedright,singlelinecheck=off]{caption}

\makeatletter
\renewcommand{\@biblabel}[1]{\quad#1.}
\makeatother

\usepackage{lastpage,fancyhdr,graphicx}
\usepackage{epstopdf}
\fancyheadoffset[L]{2.25in}
\fancyfootoffset[L]{2.25in}
\begin{document}
\vspace*{0.2in}

\begin{flushleft}
{\Large
\textbf\newline{Constrained Linear Movement Model (CALM): Simulation of passenger movement in airplanes} 
}
\newline
\\
Mehran Sadeghi Lahijani\textsuperscript{1\Yinyang*},
Tasvirul Islam\textsuperscript{2\ddag},
Ashok Srinivasan\textsuperscript{2\ddag},
Sirish Namilae\textsuperscript{3\ddag}
\\
\bigskip
\textbf{1} Department of Computer Science, Florida State University, Tallahassee, Florida, USA
\\
\textbf{2} Department of Computer Science, University of West Florida, Pensacola, Florida, USA
\\
\textbf{3} Aerospace Engineering Department, Embry-Riddle Aeronautical University, Daytona Beach, Florida, USA
\\
\bigskip

%
%
\Yinyang These authors contributed equally to this work.

\ddag These authors also contributed equally to this work.




* sadeghil@cs.fsu.edu

\end{flushleft}
\section*{Abstract}
Pedestrian dynamics models the walking movement of individuals in a crowd. It has recently been used in the analysis of procedures to reduce the risk of disease spread in airplanes, relying on the SPED model. This is a social force model inspired by molecular dynamics; pedestrians are treated as point particles, and their trajectories are determined in a simulation. A parameter sweep is performed to address uncertainties in human behavior, which requires a large number of simulations. 

The SPED model’s slow speed is a bottleneck to performing a large parameter sweep. This is a severe impediment to delivering real-time results, which are often required in the course of decision meetings, especially during emergencies. We propose a new model, called CALM, to remove this limitation. It is designed to simulate a crowd’s movement in constrained linear passageways, such as inside an aircraft. We show that CALM yields realistic results while improving performance by two orders of magnitude over the SPED model.

\section*{Introducion}

Pedestrian dynamics deals with modeling the movement of individuals, often as a part of a crowd. Its has been used in a wide variety of applications, from panic simulation and crowd behavior analysis~\cite{Helbing2009, garcimartin2017pedestrian, zou2016evacuation, Helbing-2000, Mehran-2009, Sivers-2016, moussaid2010walking} to diseases-spread modeling and robotics~\cite{Chunduri-2018, Namilae-2017b, Namilae-2017a, Sivers-2016, Song-2016, Srinivasan-2016, knepper2012pedestrian}.

The most popular models for the simulation of pedestrian dynamics are social force models~\cite{helbing1995social}. Two sets of forces determine the movement of each person in these models. Propulsive forces tend to move a pedestrian toward that pedestrian's destination. Repulsive forces, induced by nearby pedestrians or physical obstacles in the environment, on the other hand, tend to impede this movement. Newton's law of motion is used with the aggregation of the above forces to compute the trajectories of the pedestrians. These trajectories are subsequently analyzed to extract the desired information for the target application.

The Self Propelled Entity Dynamics (SPED) model  is one notable social dynamics model~\cite{Chunduri-2018,Namilae-2017b,Namilae-2017a,Srinivasan-2016}. It has been used to determine contact patterns while passengers board and disembark from planes. These contact patterns are subsequently used to identify policies that would hinder the spread of viral diseases in air-travel~\cite{Namilae-2017a}. The results of this research, analyzing the risk of an Ebola outbreak, was widely covered in more than 75 news outlets around the world~\cite{VIPRA_News}. 

A critical challenge in the use of such models, especially during epidemics, lies in dealing with the intrinsic uncertainties in human behavior. The above application handles it by parameterizing the sources of uncertainty and then performing a sweep of the parameter space to generate all possible scenarios~\cite{Srinivasan-2016,Chunduri-2018}. Vulnerabilities of different policies can then be examined under this exhaustive set of possible scenarios.

The number of scenarios that need to be generated is large, leading to a high computational effort. Chunduri et al.~\cite{Chunduri-2018} observed that it is crucial to have the results of decision-support systems in less than a couple of minutes during decision meetings. The average runtime of SPED code for a {\em single simulation} is over 350 seconds. This makes it infeasible to meet the time constraints needed for decision support. Consequently, there is a need for a model that can simulate the same application as SPED, while being faster.

The primary contribution of this paper lies in proposing a fast pedestrian dynamics model -- \textit{CALM} -- for constrained linear movement of individuals in a crowd. Similar to SPED, this model is designed to simulate  movement in narrow, linear, passageways, such as inside  airplanes. Our results show that CALM performs 72 times faster than the SPED model. Apart from this performance gain, we have modeled additional behavioral features of pedestrians. Therefore, the CALM model can overcome the limitations of SPED in a decision support context where real time results are required.

\section*{Model Description}
\label{sec:model}

Various types of methods have been used in pedestrian dynamics, such as cellular automaton~\cite{schadschneider2001cellular, burstedde2001simulation}, social force~\cite{helbing1995social, Mehran-2009}, and lattice-gas~\cite{muramatsu1999jamming,zou2016evacuation} models. Social force models are among the most popular, and are particularly effective in generating fine-scale trajectories of each pedestrian. Both the existing SPED model and the proposed CALM model are social force models. We first summarize the SPED model and then explain how its critical qualitative features are carried forward to the CALM model, while enabling it to be computationally more efficient. 

\subsection*{SPED Model}
\label{sec:sped}

Social force models treat each pedestrian as a point particle, analogous to an atom in molecular dynamics. Molecular dynamics uses models that capture the actual attractive and repulsive forces between atoms, which govern the movement of atoms. In social force models, conceptual forces are defined that perform a similar role, either increasing the speed or decreasing it. Each pedestrian wishes to reach a certain destination, which motivates a propulsive force $f^{prop}$. The pedestrian may be hindered by other pedestrians and fixed objects, whose net effect is considered a repulsion, $f^{rep}$. Newton's law of motion, shown in Equation~(\ref{eq:newton}), is then solved using a finite difference solver to determine the position $x_i$ and velocity $v_i$ of each pedestrian $i$ with mass $m_i$ as a function of time $t$.
\begin{equation}
\label{eq:newton}
    m_i\frac{dv_i}{dt} = m_i\frac{d^2x_i}{dt^2} =  f_i^{prop}(t) + f_i^{rep}(t)
\end{equation}

Both SPED and CALM use the model for propulsive force given in Equation~(\ref{eq:Propulsion}), which is common in social dynamics~\cite{helbing1995social}. Here, $\tau$ is the average reflex time of an individual  and $v_{0i}$ is the desired velocity of pedestrian $i$ (if there were no obstacles or other pedestrians). SPED assigns normally distributed random values for $|v_{0i}|$, with the direction of $v_{0i}$ determined by the pedestrian's destination. The parameters for the distribution are based on empirically observed values for human movement~\cite{zkebala2012pedestrian}.
\begin{eqnarray}
\label{eq:Propulsion}
    f^{prop}_{i} = {\frac{(v_{0i} - v_i)}{\tau} \times m_i}
\end{eqnarray}

Social dynamics models typically differ in their definition of the repulsive force. The basic idea is that the speed of a pedestrian should decrease on getting close to other persons or fixed objects on their path.

SPED computes repulsion for each pedestrian as follows. It considers the distance $d_i$ of the closest pedestrian or fixed object in the direction of motion. It then uses one of the following criteria, depending on the value of $d_i$. (i) If $d_i$ is very large, then it ignores the repulsive force in Equation~(\ref{eq:newton}). (ii) If $d_i$ is very small, then it adapts the Lennard-Jones potential from molecular dynamics to compute $f_i^{rep}$ in Equation~(\ref{eq:newton}) as explained below. (iii) If $d_i$ is between these extremes, then it decreases the speed of the pedestrian by multiplying the speed of the pedestrian by a factor $\alpha$ to obtain the speed at the next time step of the finite difference scheme. That is, $v_i(t+\Delta t) \leftarrow \alpha v_i(t)$, where $\alpha$ is computed using  Equation~(\ref{eq:alpha}). In that equation, $\lambda$ is a constant giving the desired stopping threshold of a pedestrian; that is, the pedestrian aims to stop when the distance to the nearest pedestrian reaches this value.  
\begin{eqnarray}
\label{eq:alpha}
    \alpha = { 1 - \frac{\lambda}{d_i}}
\end{eqnarray}

A modified version of the Lennard-Jones potential from molecular dynamics is used to compute repulsion when $d_i$ is very small. In that case, the distance $r_{ij}$ to each of the other pedestrians $j$ is determined and a "potential" computed for each as shown in Equation~(\ref{eq:SPED_LJ})~\cite{Namilae-2017a}. The net repulsive force on pedestrian $j$ is given by $f_i^{rep} = \sum_{j \neq i} f_{ij}$, where $f_{ij}$ is the negative of the gradient of the potential $V_{ij}^{LJ}$. Repulsion is computed in an identical manner if the closest entity is a fixed obstacle rather than a pedestrian.

\begin{eqnarray}
\label{eq:SPED_LJ}
    V^{LJ}_{ij} = {\epsilon{\frac{\sigma}{r_{ij}^{12}}}}
\end{eqnarray}
Here, $\epsilon$ and $\sigma$ are suitably defined constants obtained by fitting against empirically determined values. A few other behavioral characteristics are directly incorporated into the SPED code. (i) Passengers take a certain time to stow their luggage while boarding. (ii) Passengers take a certain time to retrieve their luggage while disembarking. (iii) Passengers let those in the row ahead of them leave first while disembarking. 

We observed by profiling the code that the Lennard-Jones potential is the primary computational bottleneck. In addition, the methodology to reduce the speed in the intermediate range is numerically awkward for the following reason. In order to verify convergence, a typical test would be to repeat the simulation with a smaller time step size and check if the results are similar. However, the method adopted in SPED makes the velocity a function of the time step size, which would not yield similar results when the time step size is changed. We designed the CALM model to overcome both limitations, and also added an additional behavioral feature that ensures that the simulation always progresses.  

\subsection*{CALM Model}
\label{sec:calm}

We note that actual human movement is not precisely defined by any particular potential. Rather, it varies so much that it is sufficient to capture its qualitative tendencies, and then use a parameter sweep to examine the range of movement patterns. We, therefore, define a simpler repulsive force by using a single curve that yields results qualitatively similar to SPED. It must satisfy the basic requirements that the speed should be the desired speed $|v_{0i}|$ when the nearest pedestrian is far away, and should decrease to 0 when the nearest passenger gets too close.

We accomplish this by using Equation~(\ref{eq:newton}) with propulsion defined by Equation~(\ref{eq:Propulsion}) as with SPED, and repulsion by Equation~(\ref{eq:CALM1}) as defined below, for a value of $\beta$ explained below.
\begin{eqnarray}
\label{eq:CALM1}
    f^{rep}_{i} = {\frac{(\beta - 1) v_{0i}}{\tau} \times m_i}
\end{eqnarray}

\begin{figure}[!h]
\includegraphics[width=1.0\linewidth]{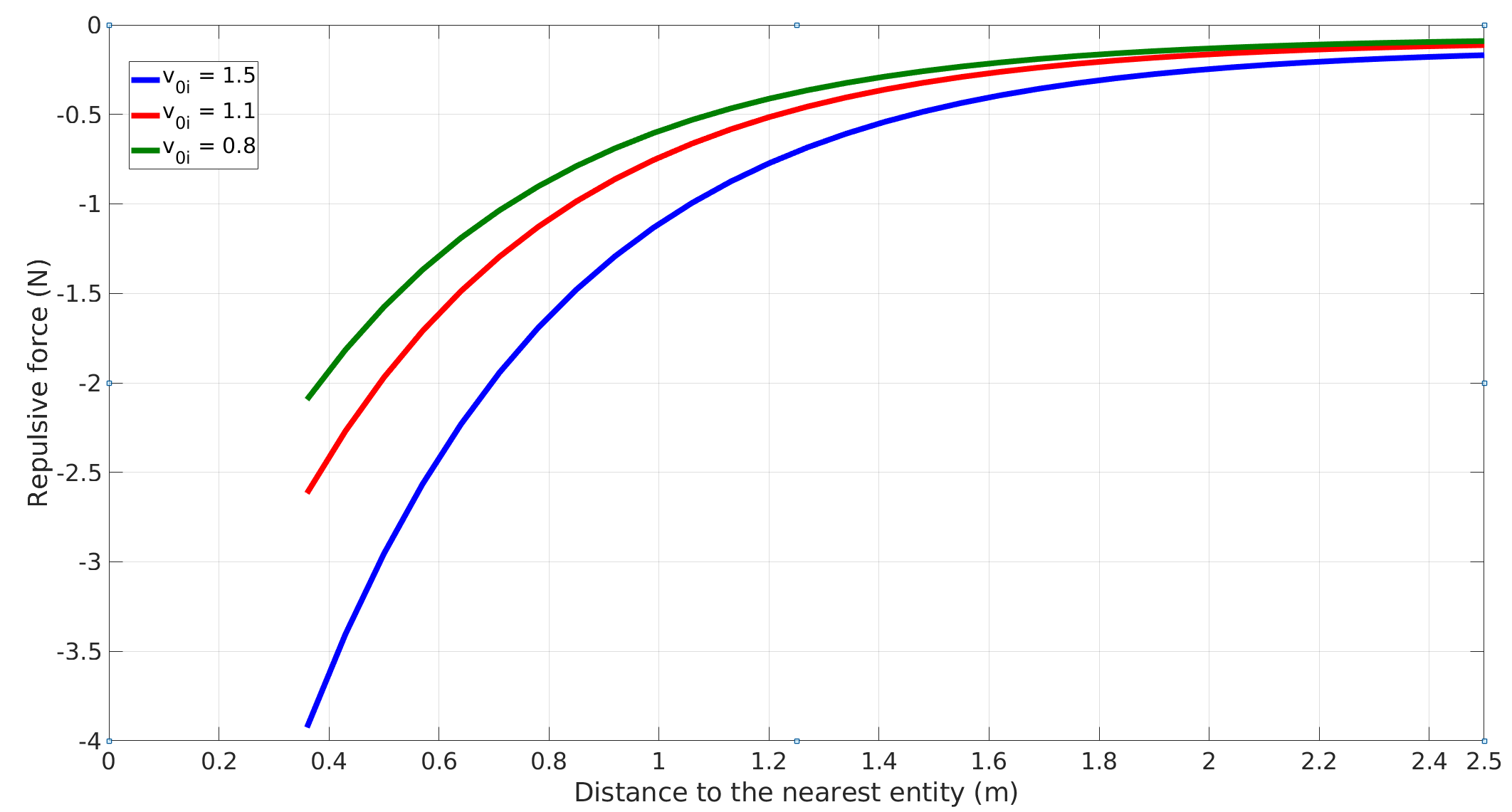}
\caption{{\bf Repulsive force as a  function of $d_i$}
}
\label{fig:CALM_Repulsion}
\end{figure}

Figure~\ref{fig:CALM_Repulsion} shows the values of this repulsion against $d_i$ with three different values of $v_{0i}$. The force becomes negative when a pedestrian reaches close to another, with stronger repulsion for faster moving pedestrians. 

Use of the above repulsive force is equivalent to solving Newton's law of motion with the net force  defined by Equation~(\ref{eq:CALM2}) below, which is the actual equation that we solve.
\begin{equation}
\label{eq:CALM2}
    m_i\frac{d^2x_i}{dt^2} =  {\frac{(\beta v_{0i} - v_i)}{\tau} \times m_i}
\end{equation}
Comparing the above with the expression in Equation~(\ref{eq:Propulsion}) shows that the above formulation aims to achieve a velocity of $\beta v_{0i}$. We choose $\beta$ as a function of $d_i$ such that $\beta v_{0i}$ is a suitable velocity for the pedestrian whose nearest pedestrian is at a distance $d_i$.

\begin{figure}[!h]
\includegraphics[width=1.0\linewidth]{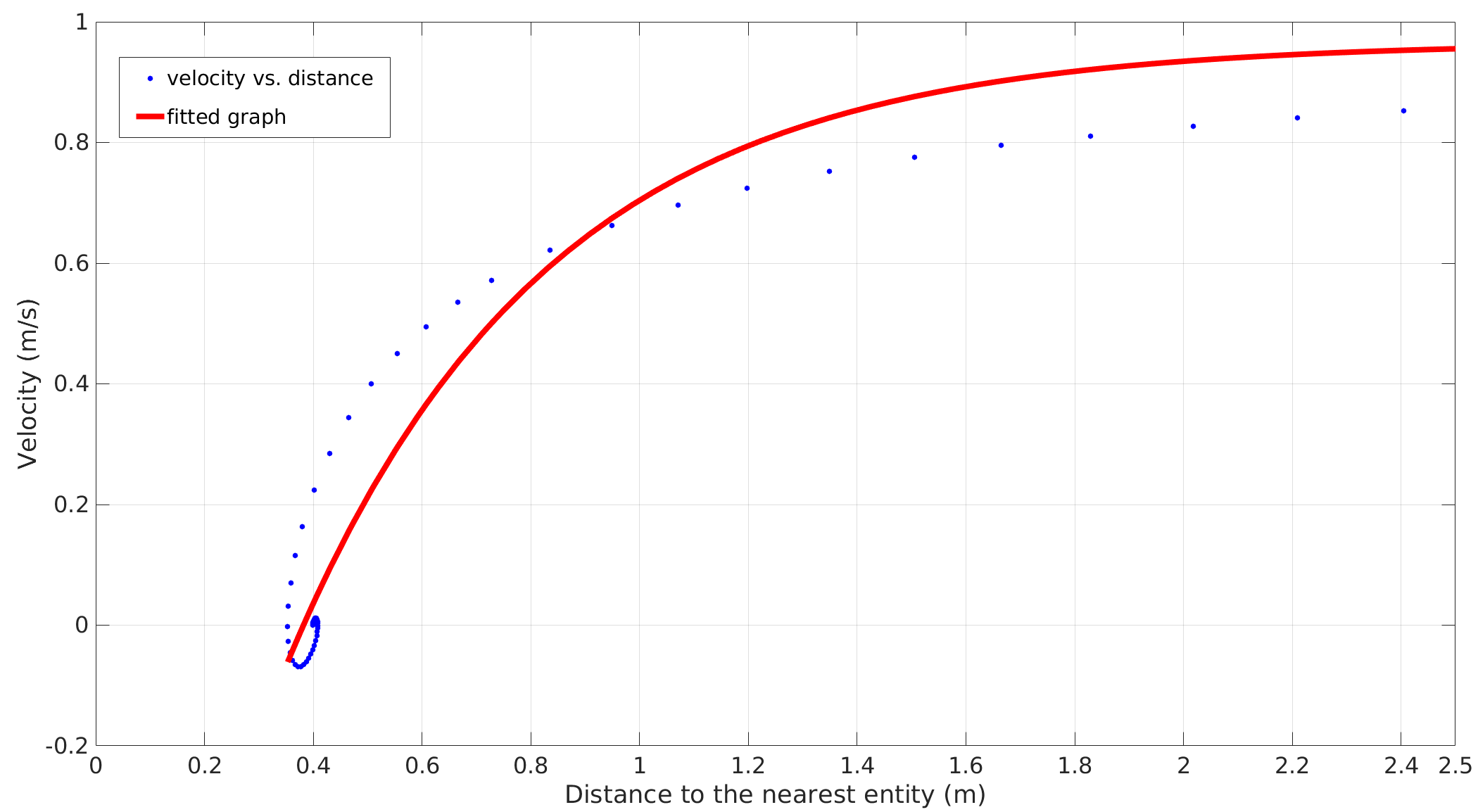}
\caption{{\bf Fitting the repulsion term of CALM (solid line) to the SPED solution (dotted line).}
}
\label{fig:fittedGraph}
\end{figure}

We accomplish this goal as follows. We calculate the variation of $v_i$ with $d_i$ for SPED and fit a curve that will yield a qualitatively similar result for CALM. The specific details are as follows. We note that the typical desired speed $v_{0i}$ varies around 1m/s. We solve SPED's equation when there are only two pedestrians, one stationary and another moving toward that pedestrian with an initial velocity of 1m/s while at a distance of 4m, with only the repulsion term considered. The intermediate zone is taken as a distance from 0.4m to 2.0m. We get the result shown in Figure~\ref{fig:fittedGraph}.

In fitting the CALM model, we assume $\beta$ of the form given in Equation~(\ref{eq:CALM_DS}) and determine parameters that give a best fit to the SPED results. We obtain $a = 2.11$, $b = 0.366$, and $c = 0.966$. We use the same value of $\beta$ for any value of $v_{0i}$. The range of values of $v_{0i}$ is small around 1ms/, and so the same value of $\beta$ is considered a reasonable scaling factor. 
\begin{eqnarray}
\label{eq:CALM_DS}
    \beta = {c - e^{-a(d_i - b)}}
\end{eqnarray}

\begin{figure}[!h]
\includegraphics[width=1.0\linewidth]{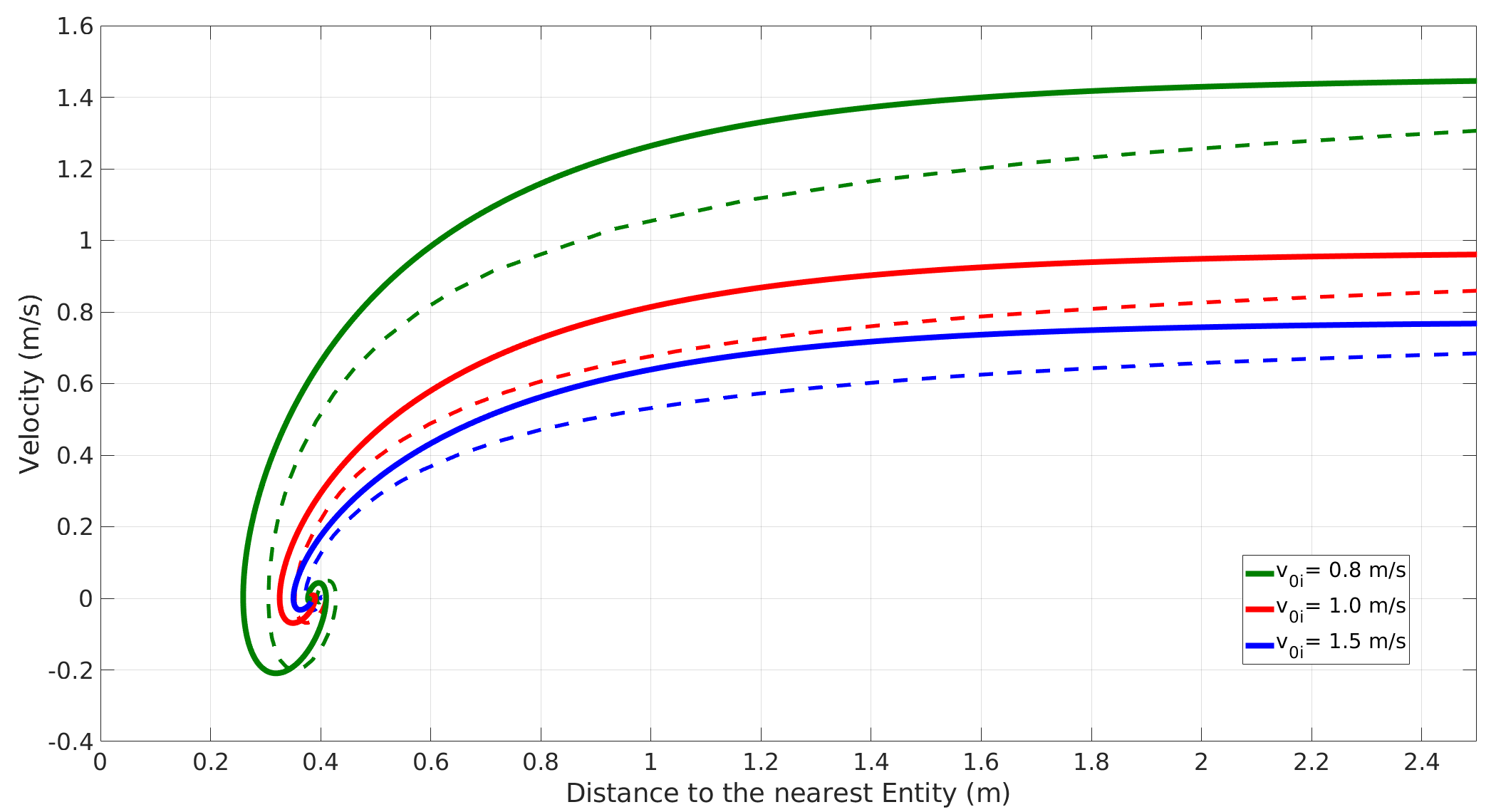}
\caption{{\bf Comparing the pedestrian dynamics results of the SPED model (dashed line) vs the CALM model (contiguous line) in three different cases}
}
\label{fig:CALMvsSPED}
\end{figure}

Figure~\ref{fig:CALMvsSPED} demonstrates a comparison of the pedestrian dynamics results between the CALM and the SPED model in three different cases of $v_{0i} = 0.8 (m/s)$ , $v_{0i} = 1.0 (m/s)$, $v_{0i} = 1.5 (m/s)$. In each case, we assumed that a pedestrian is moving toward a stationary entity. These figures show a qualitative match between SPED and our model for this situation. {\em Given a range of values of $v_{0i}$ for SPED, we can find a range of values for CALM that yield similar results}.

\paragraph{Deadlock} We incorporate the same behavioral features as SPED in our simulations. In addition, we include the following. We have designed the CALM model for simulation of movement in narrow passageways, such as airplane aisles. In these situations, people cannot walk side-by-side. As a consequence, sometimes, a deadlock situation can happen when two people try to get right of way but neither can move forward. The SPED model will not progress in these situations. 

We have designed a mechanism to resolve this. One of the passengers will be declared the winner, with a random component to the decision, and get the right of way. This reflects human behavior in practice, where one person would yield to another.  

\section*{Application to Air Travel}

 As we discussed in the introduction section, using pedestrian dynamics models for simulation of passengers movement in the airplanes is a critically important application for public health policy analysis. Namilae et al.~\cite{Namilae-2017a} used the SPED model for suggesting policies that help to mitigate the risk of diseases spread during the flights. We use the CALM model in the same application to demonstrate that it yields realistic results. In this section, we explain the implementation details for airplane boarding and deplaning procedures.

\subsection*{Deplaning}
\label{sec:deplaning}

We initialize the simulation by inputting the initial positions of passengers and physical obstacles from a file. We also assign each passenger a value of $|v_{0i}|$ drawn independently from a Gaussian distribution with mean $\overline{v_0}$ and standard deviation 0.2 m/s. The parameter $\overline{v_0}$ varies in different simulations, with $1.1\leq\overline{v_{0}}\leq1.3$ m/s. These values are derived from empirical data available for pedestrian speed~\cite{zkebala2012pedestrian}.

After the initialization, each step of the simulation computes the position of all passengers at several different points in time. This is accomplished by using an explicit Euler scheme to solve Equation~(\ref{eq:newton}) with a time-step size $\Delta$t of $0.005$s for the CALM model. We repeat these steps until all the passengers leave the plane. Algorithm 1 gives the pseudocode of the deplaning procedure.

\begin{algorithm}
\caption{Deplaning}
	\begin{algorithmic}

   	\WHILE {there are passengers in the plane}
   		\FOR{each remaining passenger $P_i$}
   				\STATE Find the nearest passenger or physical obstacle to $P_i$ on its path
   				\STATE Compute the repulsion
   		\ENDFOR
   		\FOR{each remaining passenger $P_i$}
   					\STATE Compute the propulsion 
   					\STATE Update the velocity and position of $P_i$
		    		\STATE Check for deadlock and update $P_i$'s state if necessary
   		\ENDFOR
   	\ENDWHILE
	\end{algorithmic}
\end{algorithm}

\begin{figure}[!h]
\includegraphics[width=1.0\linewidth]{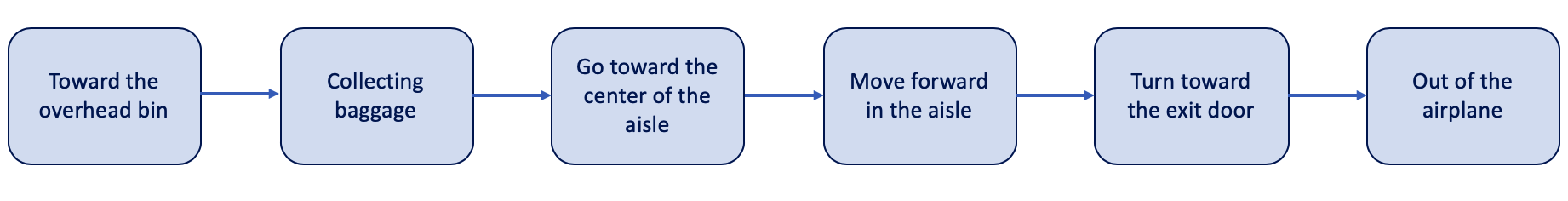}
\caption{{\bf States of passengers while deplaning}
}
\label{deplaningSD}
\end{figure}

\begin{figure}[!h]
\includegraphics[width=1.0\linewidth]{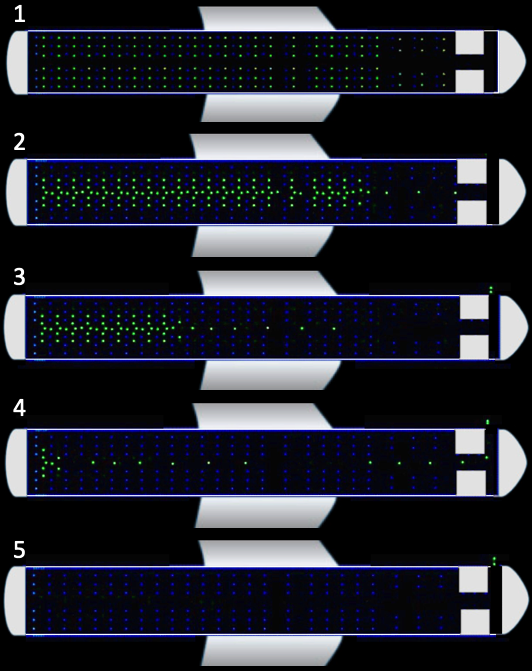}
\caption{{\bf Progress of a deplaning simulation}
}
\label{photoset}
\end{figure}

During the deplaning procedure, each passenger will go through a few different states. The initial state of each passenger is going {\em toward the overhead bin} to collect the carry-on baggage. We assume that each passenger has a bag that is in the nearest overhead bin and that they take 5-12s in a state where they are {\em collecting baggage}. Passengers will then attempt to {\em go toward the center of the aisle} and subsequently {\em move forward in the aisle} toward the exit. In the latter state, they will first wait for the passengers in the rows in front of them to go ahead before they move forward. When passengers reach the end of the aisle, they have to {\em turn toward the exit door} and leave the plane. Once a passenger is {\em out of the airplane}, we remove that passenger from the simulation. Figure~\ref{deplaningSD} shows the state diagram for passengers during the deplaning, and Figure~\ref{photoset} demonstrates the progress of a disembarkation simulation.

In addition to $\overline{v_0}$, we use five parameters to address the uncertainties in human behavior during the deplaning process. Passengers move from the seat to the overhead bin at a fraction of their maximum reachable speed. Therefore, in our model, we use a coefficient $0<toward\_bag\_speed\_coefficient<1$ that multiplies $v_{0i}$ when passengers are in the $toward\_the\_overhead\_bin$ state. Similarly, passengers’ alignment on the center of the aisle, after taking their bag, takes place with a fraction $0<aligning\_speed\_coefficient<1$ of their maximum reachable speed. After this step, passengers should wait for the passengers in the front rows to move toward the end of the aisle first. Hence, the first passenger from row $i$ can proceed only if the last passenger from row $i-1$ has already moved in the aisle for $aisle\_distance\_threshold$ meters. People often reduce their speed when getting close to an intersection at which they want to turn. In our model, passengers, reduce their speed by a factor $intersection\_speed\_coefficient$ when their distance to the end of the aisle is less than $intersection\_distance\_threshold$ meters.

According to~\cite {wald2014structured}, the average deplaning rate for the planes is 15-17 passengers per minute. In our application, the goal is to capture some extreme scenarios in addition to normal ones~\cite{Chunduri-2018}. As a consequence, we tuned the parameter values for the CALM model to generate simulations with a slightly more extensive range of deplaning rates. We use 1000 parameter combinations for running simulations of the deplaning procedure for a full Airbus A320 with 144 seats and used the results of this parameter sweep for tuning the parameters of the CALM model. For this airplane, the deplaning rate of 15-17 passengers per minute yields to deplaning time of 8.47 to 9.6 minutes, and the CALM model generates simulations with the deplaning time of 6.52 to 11.10 minutes after parameter tuning. Table~\ref{deplaningParams} shows the range of values for all six parameters of the CALM model after such adjustment.

\begin{table}[!ht]
\centering
\caption{
{\bf CALM model parameter ranges for deplaning}}
\begin{tabular}{|l|l|l|}
\hline

{\bf Parameter} & {\bf Minimum Value} & {\bf Maximum Value} \\ \thickhline
$\overline{v_{0}}$ & 1.1 m/s & 1.3 m/s \\ \hline
$toward\_bag\_speed\_coefficient$ & 0.2 & 0.6 \\ \hline
$aligning\_speed\_coefficient$ & 0.2 & 0.7 \\ \hline
$aisle\_distance\_threshold$ &  0.5 m & 1.6 m \\ \hline
$intersection\_speed\_coefficient$ & 0.2 & 0.8 \\ \hline
$intersection\_distance\_threshold$ & 0.2 m & 1.5 m \\ \hline

\end{tabular}
\label{deplaningParams}
\end{table}

\subsection*{Boarding}

\begin{figure}[!h]
\includegraphics[width=1.0\linewidth]{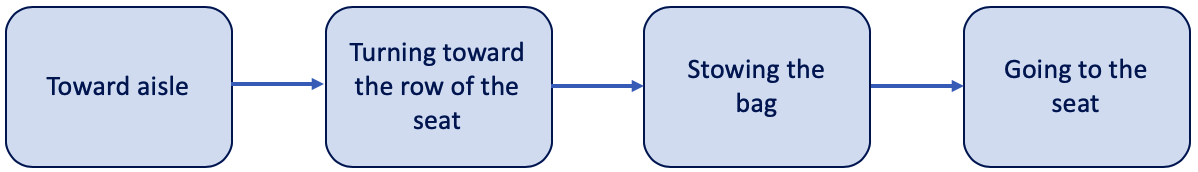}
\caption{{\bf State diagram of passengers during the boarding procedure}
}
\label{boardingSD}
\end{figure}

The CALM model can also be used for boarding of airplanes with implementation details being similar to deplaning to a large extent. The main difference lies in the state diagram of the passengers, which is roughly in the reverse order of deplaning, as shown in Figure~\ref{boardingSD}. Passengers are initially outside of the airplane. As the simulation starts, passengers move toward the aisle from the exit door. Then, they go into the aisle until they reach the row of their seat. There, they will stow their bag in the overhead bin, and then go to their seat. 
 
 We have implemented boarding with three zones. Other boarding strategies can be implemented without changing the code, just by using suitable input files.

Boarding involves a few parameters, similar to deplaning. The parameter $\overline v_{0}$  was explained earlier. We assume passengers are in an aisle outside the airplane (behind the airplane door) and will move toward the airplane door when the passenger in front of them has moved for $line\_distance\_threshold$ meters. Passengers reduce their speed by a factor $intersection\_speed\_coefficient$ when their distance to the intersection of the aisle is less than $intersection\_distance\_threshold$ meters. After passengers place their baggage in the overhead bin, they will move toward their seat with a fraction of their maximum reachable speed, that we call $toward\_seat\_speed\_coefficient$. These parameters are analogous to those for deplaning, and we use the same range of values for them, as shown in Table~\ref{boardingParams}.

\begin{table}[!ht]
\centering
\caption{
{\bf CALM model parameter ranges for boarding}}
\begin{tabular}{|l|l|l|}
\hline

{\bf Parameter} & {\bf Minimum Value} & {\bf Maximum Value} \\ \thickhline
$\overline{v_{0}}$ & 1.1 m/s & 1.3 m/s \\ \hline
$line\_distance\_threshold$ & 0.5 m & 1.6 m \\ \hline
$intersection\_speed\_coefficient$ & 0.2 & 0.8 \\ \hline
$intersection\_distance\_threshold$ & 0.2 m & 1.5 m \\ \hline
$toward\_seat\_speed\_coefficient$ & 0.2 & 0.6 \\ \hline

\end{tabular}
\label{boardingParams}
\end{table}

\section*{Results}

We first show that the CALM models results are consistent with empirically observed metrics for disembarkation in airplanes. We then show that the CALM model yields substantial performance gains over the SPED model. 

\subsection*{Experimental Setup}

We run all our experiments on the Frontera supercomputer at the Texas Advanced Computing Center. This system consists of 8008 compute nodes with 56 cores per node for a total of 448448 cores and ranks the 5th fastest supercomputer in the world. Each node contains two Xeon Platinum 8280 28C processors running at 2.7GHz with 128 GB memory. Nodes are connected through Mellanox Infiniband HDR-100 network connected in a fat tree topology. 

We use a scrambled Halton low dispcrepancy sequence for performing efficient parameter sweep~\cite{Chunduri-2018}. We employ dynamic load balancing for running this parameter sweep, using a master-worker algorithm where the master assigns a simulation to a core that has just completed its previous simulation. We use one node with 32 cores in all of our experiments. Each simulation of the parameter sweep is run on one core, with each core running multiple simulations. 

Implementations of the CALM model for running parameter sweeps of deplaning and boarding simulations can be found at \href{https://gitlab.com/Mehran_SL/calm}{https://gitlab.com/Mehran\_SL/calm}.

\subsection*{Model Validation}

We validate our results by examining disembarkation times on three different types of airplanes. A single simulation does not capture the variety of human movement patterns, and so we perform a parameter sweep with 1000 different combinations of parameter values, covering the range mentioned in Table~\ref{deplaningParams}. We compare it against the empirically observed deplaning rate of 15-17 passengers per minute~\cite {wald2014structured}. The airplanes considered are: Boeing B757-200 with 182 seats, B757-200 with 201 seats and CRJ-200 with 50 seats. In all of our experiments, all the planes are full.  

Table~\ref{tableDeplaningResults} shows the expected ranges for deplaning time of each airplane along with the results from our simulations using the CALM model.

\begin{table}[!ht]
\centering
\caption{
{\bf Comparison of predicted and empirically observed deplaning times}}
\begin{tabular}{|l|l|l|}
\hline

{\bf Airplane} & {\bf Deplaning time} & {\bf Expected deplaning time} \\ \thickhline
$B757-200\ (182\ seats)$ & [8.21, 16.43] (min) & [10.71, 12.13] (min) \\ \hline
$B757-200\ (201\ seats)$ & [9.32, 15.17] (min) & [11.82, 13.4] (min) \\ \hline
$CRJ-200\ (50\ seats)$ & [1.46, 4.06] (min) & [2.94, 3.33] (min) \\ \hline

\end{tabular}

\label{tableDeplaningResults}
\end{table}

As the results of our experiments show, the expected deplaning time ranges are subsets of the ranges produced by the CALM model. We can draw two conclusions from these results. First, the results demonstrate that the CALM model generates results for all the expected scenarios. Second, the CALM model provides results that are outside the normal range. This was a deliberate design choice in selecting the parameter range because our application goal is to generate rare scenarios that can capture extreme events~\cite{Chunduri-2018}.

In addition to validating the deplaning times, we selected several random simulations and checked the video output to examine if the behavior was realistic, as is commonly done in validation of pedestrian dynamics~\cite{namilae2017multiscale}. Simulation results in all these tests were consistent with empirically observed behavior of passengers during deplaning and boarding. 

\subsection*{Performance Analysis}

We compare the performance of CALM and SPED using a  parameter sweep of size 1000 for disembarkation process on an Airbus A320 with 144 seats. The results show that the average runtime for simulations of the SPED model is 277.52 seconds while it is only 3.86 seconds for simulations of the CALM model. Consequently, the CALM model performs approximately 72 times faster than the SPED model. The parameter sweep of the SPED model took 9207.6 seconds while the parameter sweep of the CALM model took 128.6 seconds to complete. Therefore, the CALM model runs approximately 70 times faster than the SPED model when used for the parameter sweep of size 1000. 

There are two significant reasons for this considerable performance difference between these two models. First, we used a simpler force formulation to decrease its computational time in our model. In particular, much of the reduction in time was obtained by eliminating the Lennard-Jones potential and using a single simple formula to determine the impact of repulsion. This was the primary reason for decrease in simulation time, and led to around a factor 20 improvement in performance. Second, we removed passengers that had reached their destinations from the simulation. Both SPED and CALM models use algorithms of $O(N^2)$ time complexity where N is the number of passengers in the simulation. For the SPED model, this N is constant during the whole simulation. By removing passengers, N linearly decreased in the CALM model. This provided the secondary performance boost, yielding a factor 3.5 improvement.

\section*{Conclusions}

 Pedestrian dynamics is finding increased applications in diverse real-world problems. However, slow performance of existing models has been a bottleneck for policy analysis, especially in an emergency. In decision support meetings, a result is required in the order of a couple of minutes. The new CALM model delivers results in that time frame in contrast to the SPED model. Our validation simulations also show that the CALM model produces results that are consistent with empirical observations.

\section*{Acknowledgments}
This material is based upon work supported by the National Science Foundation under Grant No. 1640822. Also, the authors acknowledge the Texas Advanced Computing Center (TACC) at The University of Texas at Austin for providing HPC resources that have contributed to the research results reported within this paper. URL: http://www.tacc.utexas.edu.\\
The authors would like to acknowledge Robert Pahle for providing a webservice for producing the video outputs of the simulations. They would also like to thank Pierrot Derjany for helpful discussions about the SPED model.

%
%
%





\end{document}